\documentclass[sigconf]{acmart}
\AtBeginDocument{%
  \providecommand\BibTeX{{%
    \normalfont B\kern-0.5em{\scshape i\kern-0.25em b}\kern-0.8em\TeX}}}

\usepackage{stfloats}
\usepackage{graphics}
\usepackage{url}
\usepackage{hyperref}
\usepackage[normalem]{ulem}

\copyrightyear{2022}
\acmYear{2022}
\setcopyright{acmlicensed}\acmConference[CHI '22]{CHI Conference on Human
Factors in Computing Systems}{April 29-May 5, 2022}{New Orleans, LA, USA}
\acmBooktitle{CHI Conference on Human Factors in Computing Systems (CHI
'22), April 29-May 5, 2022, New Orleans, LA, USA}
\acmPrice{15.00}
\acmDOI{10.1145/3491102.3517698}
\acmISBN{978-1-4503-9157-3/22/04}

\newcommand{\name}{FaceOri}
\newcommand \change[1]{{\textcolor{black}{#1}}}
\newcommand \del[1]{}

\begin{document}


\title[\name{}: Tracking Head Position and Orientation...]{\name{}: Tracking Head Position and Orientation Using Ultrasonic Ranging on Earphones}

\author{Yuntao Wang}
\authornote{The authors contribute equally to this paper.}
\email{yuntaowang@tsinghua.edu.cn}
\affiliation{%
  \department{Key Laboratory of Pervasive Computing, Ministry of Education, Department of Computer Science and Technology}
  \institution{Tsinghua University}
  \city{Beijing}
  \country{China}
  \postcode{100084}
}

\author{Jiexin Ding}
\authornotemark[1]
\email{djx031906@163.com}
\affiliation{%
    \department{Global Innovation Exchange (GIX) Institute}
  \institution{Tsinghua University}
  \city{Beijing}
  \country{China}
}

\author{Ishan Chatterjee}
\email{ishan.chatterjee1@gmail.com}
\affiliation{%
    \department{Paul G. Allen School of Computer Science and Engineering}
   \institution{University of Washington}
  \city{Seattle}
  \state{WA}
  \country{USA}
}

\author{Farshid Salemi Parizi}
\email{farshid@uw.edu}
\affiliation{%
    \department{Electrical and Computer Engineering}
   \institution{University of Washington}
  \city{Seattle}
  \state{WA}
  \country{USA}
}

\author{Yuzhou Zhuang}
\email{zhuangyz19@mails.tsinghua.edu.cn}
\affiliation{%
    \department{Global Innovation Exchange (GIX) Institute}
  \institution{Tsinghua University}
  \city{Beijing}
  \country{China}
}

\author{Yukang Yan}
\email{yyk@mail.tsinghua.edu.cn}
\affiliation{%
    \department{Department of Computer Science and Technology}
  \institution{Tsinghua University}
  \city{Beijing}
  \country{China}
}
\authornote{denotes as the corresponding author.}

\author{Shwetak Patel}
\email{shwetak@cs.washington.edu}
\affiliation{%
  \department{Paul G. Allen School of Computer Science and Engineering}
  \institution{University of Washington}
  \city{Seattle}
  \state{WA}
  \country{USA}
}

\author{Yuanchun Shi}
\email{shiyc@tsinghua.edu.cn}
\affiliation{%
  \department{Department of Computer Science and Technology}
  \institution{Tsinghua University}
  \city{Beijing}
  \postcode{100084}
  \country{China}
}

\renewcommand{\shortauthors}{Yuntao Wang, et al.}


\begin{abstract}
Face orientation can often indicate users’ intended interaction target. In this paper, we propose \name, a novel face tracking technique based on acoustic ranging using earphones.
\name{} can leverage the speaker on \del{any}\change{a} commodity device to emit an ultrasonic chirp, which is picked up by the set of microphones on the user's earphone, and then processed to calculate the distance from each microphone to the device.
These measurements are used to derive the user’s face orientation and distance with respect to the device.
We conduct a ground truth comparison and user study to evaluate \name{}'s performance.
The results show that the system can determine whether the user orients to the device at a 93.5\% accuracy within a 1.5 meters range.
Furthermore, \name{} can continuously track user's head orientation with a median absolute error of 10.9 mm in the distance, 3.7$^\circ$ in yaw, and 5.8$^\circ$ in pitch.
\name{} can allow for convenient hands-free control of devices and produce more intelligent \change{context-aware interactions.} 

\end{abstract}



\maketitle

\section{Introduction}
\label{sec:intro}
Earphones are one of the most ubiquitous wireless accessories. As a greater number of smartphones continue to drop the earphone jack, the popularity of these mobile audio devices continues to grow. With the \change{earphones}' cord getting cut, the input microphone has now migrated from a placement inline with the cable to a position at each of the user's ears. While most headsets leverage the microphone to take calls and, more recently, to enable \change{the} active noise cancellation (ANC) functionality, we find that this unique placement of these sensors can be used to unlock a broader range of context-aware interactions when the \change{earphone} is transformed into a \textit{spatial input device} via ultrasonic ranging.

\begin{figure}
  \centering
  \includegraphics[width=0.9\columnwidth]{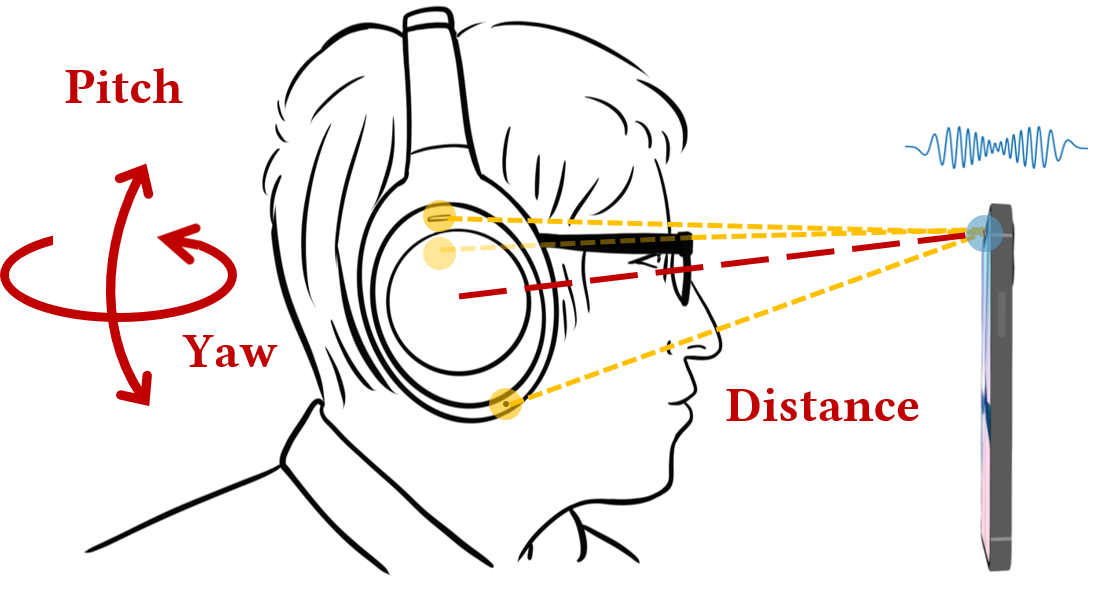}
  \caption{\name{}~tracks user's face orientation towards the device with acoustic ranging using microphones in an earphone.}
  \label{fig:intro} 
\end{figure}

In particular, users tend to orient their heads toward their intended interaction targets~\cite{attention_head}. To recognize user interaction intention, researchers have leveraged eye-gaze or head tracking by using dedicated devices~\cite{drewes2007eye,kumar2007eyepoint} or by using an external camera ~\cite{Mihai-2019-eye-contact,Mayer-worldgaze, OpenPose}. These methods apply vision-based gaze or head tracking, which carry privacy concerns~\cite{katsini2020role}, and require the user to be in the field of view of the front-facing camera.

Enabling a device to detect this intention precisely and naturally can simplify user interface flow, enable hands-free interaction, and adapt interfaces to the context of use. 
For example, smartphones can detect users' proximity and face orientation toward the device to turn on the screen and allow them to read their notifications in a hand-free manner.
Additionally, a system can accurately classify whether a user's head is oriented toward a specific device. This device-specific \textit{binary attention detector} can be used to drive context-aware experiences.
In this way, devices can adapt the layout and format of visual content based on whether the user is looking at them.
Lastly, with \textit{continuous tracking} of user proximity and face orientation, an additional set of applications including activity tracking and head gesture recognition can be realized. 

We propose \name{}, a novel interaction technique that leverages the built-in set of microphones found in almost all modern active noise canceling (ANC) earphones to infer the user’s spatial location and head orientation with respect to a smartphone, laptop, smart speaker, or other devices with a built-in speaker.
\name{} works as follows: the computing device (e.g., a smartphone) emits an inaudible, ultrasonic sound from its speaker, and the embedded microphones on the earphone receive the sound. \name{} calculates the time-of-arrival to estimate the distance from each \change{microphone} to the device even when the head occludes direct line-of-sight or introduces the Doppler effect.
\name{} uses these measurements to estimate 2-degree-of-freedom (DoF) face orientation --- pitch and yaw --- and 1-DoF distance measurement with respect to the device (Fig.~\ref{fig:intro}). The user evaluation demonstrates 93.5\% accuracy in \textit{binarized attention detection} and dynamic \textit{continuous tracking} performance to be 10.9 mm in \change{the} distance, 3.7$^\circ$ in yaw, and 5.8$^\circ$ in pitch, which significantly outperforms the baseline acoustic ranging method (CAT~\cite{Mao-MobiCom-CAT}: 42.0 mm, 11.0$^\circ$, and 11.6$^\circ$) on our collected dataset. These outputs enable many hands-free device interactions, including \change{convenient wake-up of the devices, attentive user interfaces, and fitness tracking.} To our best knowledge, we are the first to benchmark the head orientation tracking performance with acoustic ranging methods \change{using build-in microphones in commodity ANC earphones}. In this paper, we offer three main contributions:

\begin{enumerate}

\item A spatial input technique that applies ultrasonic ranging to enable continuous head orientation and distance tracking with respect to a device with a speaker \change{using a built-in set of microphones in the commodity ANC earphone}.
\item \change{An end-to-end system characterization and user evaluation demonstrate \name{}'s high dynamic performance in continuous tracking and binarized attention detection.}
\item An exploration of the application space afforded by \name{} with prototypes of selected demonstrative experiences,  showcasing the applicability of the proposed approach. 
\end{enumerate}

\section{Related Work}
\label{sec:related}
\name{} employs acoustic \change{ranging} to track the earphone's position relative to a device with a speaker (e.g., phone), enabling natural and precise face orientation based interactions. In this section, we first position this paper with respect to the attentive user interface literature. We then review the related works on acoustic \change{ranging} with a focus on mobile systems. 

\subsection{Attentive User Interfaces}

Attentive user interfaces have been proposed as a natural user interface concept, sensitive to the user’s focus of attention~\cite{Roel-AUI, AUIframework}. Gaze pointing, as one of the important input modalities, has traditionally used dedicated camera-based eye tracking technology to identify which object a person is looking at~\cite{kumar2007eyepoint, drewes2007eye, esteves2015orbits, eyepattern}. 
Researchers have explored gaze-aware solutions that enable users to start conversations with software agents~\cite{looktotalk, EyePliances}, select applications on computers~\cite{jacob, MediaEyePliances, gaze-voiceincar}, and control home appliances ~\cite{auralamp} by looking to the targets. 
However, these techniques require users to wear intrusive gaze trackers or environments to be instrumented with dedicated cameras, limiting these methods' ubiquity.

Face orientation can also be used as a proxy for \change{the user's attention}~\cite{attention_head, bakx2003facial,kinya1983effect,looktotalk,stiefelhagen2002head, headbang, Xu-2020, Yan-HeadCross}. Therefore, prior research has explored tracking users' face orientation to infer their focus toward targets within graphical user interfaces~\cite{malkewitz1998head,ronzhin2005assistive}, user authentication~\cite{Liang-AuthTrack}, smart home appliances~\cite{itoh2001multi, jeet2015radio, segura2007multimodal}, VR and AR targets~\cite{Pinpointing, Telekinesis,3D_User_Interfaces,Yan-HeadCross}, wearable computing~\cite{Multimodal_wearable} and assistive interfaces~\cite{Head_Pointing}. In industry, several different smartphone applications have been released that incorporate face tracking via the front-facing camera for experiences like Animoji, Memoji, and face filters~\cite{ARKit, ARCore, Snap}.
\change{Recent works have adopted RGB~\cite{Mihai-2019-eye-contact, Abate-2019-head-pose, fast_face_ori} or depth camera~\cite{Borghi-2020-face-from-depth, Mayer-worldgaze} to accurately track the user's head pose or gaze.}
These methods apply vision-based gaze or head tracking, which carry privacy concerns~\cite{katsini2020role}, and require the user to be in the field of view of the front-facing camera. \change{As a result, they would be incompatible} for devices without a camera, such as smartwatches~\cite{Sun-Float} or smart speakers.

There have also been related works on face orientation \change{detection} that use microphone arrays distributed around the room to predict the direction of the user's voice~\cite{orientation_acoustic_source, sound_source_orientation, DoV, jackieSoudr}. \change {Although these voice-based face orientation detection methods are wearable-free, they require users to speak to the targets. Instead, \name{} can continuously track the user's face orientation and relative distance without the requirement of speaking. Thus, \name{} can benefit a wider range of interaction scenarios (e.g., working environment). Further, \name{} has the potential to achieve higher degrees of freedom and face orientation detection performance.}

\subsection{Background on Acoustic Ranging}
Acoustic signals have been studied extensively for various tracking applications. Traditionally, acoustic tracking systems are based on the Doppler effect, which calculates the frequency shift to infer a moving object's speed, and thus distance~\cite{kalgaonkar2009one, gupta2012soundwave}.
AAMouse~\cite{yun2015turning} uses the frequency shifts of transmitted signals to enable accurate device tracking and achieves a median error of 1.4 cm. Another set of acoustic tracking systems performs auto-correlation to determine the travel time (and thus distance) between the speaker and the microphone~\cite{peng2007beepbeep, nandakumar2016fingerio}, \change{achieving centimeter-level accuracy}. Phase-based methods treat received signals as phase-modulated signals and analyze phase changes to obtain fine-grained distance information~\cite{wang2016device, zhang2017soundtrak}, achieving a mean distance error of 1.3 cm in 3D space. \change{EarphoneTrack~\cite{Cao-EarphoneTrack} adopts the speaker in the earphone as the transmitter for acoustic ranging. It utilizes the leakage signal from the earphone's speaker to the microphone as a reference signal to calculate the distance from the earphone to the connected device.} 

Most similar to our work, acoustic ranging via Frequency Modulated Continuous Wave (FMCW) was proposed for high-precision distance estimation. CAT~\cite{Mao-MobiCom-CAT} proposed a distributed FMCW technique to accurately estimate the absolute distance between a transmitter and a receiver. It further combines IMU measurements to achieve mm-level tracking performance. Based on CAT, MilliSonic~\cite{Wang-2019-MilliSonic} utilizes the phase information in the demodulated FMCW signal to compute distances and further refine the tracking accuracy. The paper prototypes a 4-microphone array setup and achieves 2.6 mm median 3D tracking accuracy for smartphones. DroneTrack~\cite{DroneTrack} applies MUltiple SIgnal Classification (MUSIC) for solving the multipath and strong noise 
issues, achieving 2-3 cm distance median error and 1$^\circ$-3$^\circ$ orientation median error.

However, these acoustic ranging methods rely on a direct line-of-sight (LOS) and low moving speed between the speaker and the microphone, limiting the applicability of the approaches to our face orientation application. In our scenario, the head occludes the direct path from the microphone to the device speaker, resulting in a severe non-LOS issue. Further, the quick head movement introduces a significant Doppler effect. These issues significantly damage the tracking performance. Inspired by advanced techniques in the FMCW radar research field~\cite{merrill2001introduction,Liang2009,Chin-NLOS}, we extended CAT~\cite{Mao-MobiCom-CAT} with optimization approaches, \change{including adopting the triangular modulated chirp signal to reduce the Doppler effect~\cite{merrill2001introduction}, and applying an advanced filtering method to increase the signal-to-noise-ratio (SNR)~\cite{richards2014fundamentals, Liang2009,Chin-NLOS}.} To our best knowledge, we are the first to introduce acoustic ranging for head orientation to a practical usage setup – commodity ANC earphone and a smart device and benchmark its performance.

\section{Method}
\label{sec:method}

\name{} tracks head \textit{distance} and head \textit{orientation} in relation to a device to feed smarter device interactions, using the following two-step process. \del{To derive estimates for these values, a two-step process is followed.} As Fig.~\ref{fig:method}A shows, a set of distance measurements from the device speaker to the earphone microphones are produced via FMCW acoustic ranging, requiring a low-effort calibration procedure (described below). Second, these distances are fed to a geometric model to continuously calculate the face orientation (both yaw and pitch) with respect to the speaker.
Separately, to enable context-aware applications that only require information on whether the user's face \change{orients} to the device or not (\textit{binarized attention detection}), we employ a binary classifier on a set of acoustic features (Fig.~\ref{fig:method}B). Notably, \textit{binarized attention detection} is calibration-free.
We describe the methods and algorithms below. 

\subsection{Acoustic Ranging Using FMCW}
\label{sec:distance}
\subsubsection{CAT Acoustic Ranging}
This section provides a brief review of the fundamental aspects of CAT~\cite{Mao-MobiCom-CAT} for our method.
The speaker emits a chirp signal whose frequency changes linearly with time, $f(t)=f_0 + \frac{B}{T}t$, where $B$ is the frequency bandwidth and $T$ is the sweep time. By integrating the frequency, we can express the transmitted signal in time domain as $y_t(t)=A_0cos(2\pi f_0t + \pi \frac{B}{T}t^2)$. After some time delay $t_d$, the microphone receives the signal as $y_r(t)=A_1cos(2\pi f_0(t-t_d)\pm\pi \frac{B}{T}(t-t_d)^2)$. By mixing the received signal with the transmitted signal and applying a low pass filter, we obtain following signal:

\begin{equation}
    y_m=\frac{A_0A_1}{2}cos(2\pi \frac{B}{T}t_dt+2\pi f_0 t_d - \pi \frac{B}{T}t_d^2)
    \label{equ:ym}
\end{equation}

\begin{figure}
  \centering
  \includegraphics[width=0.95\columnwidth]{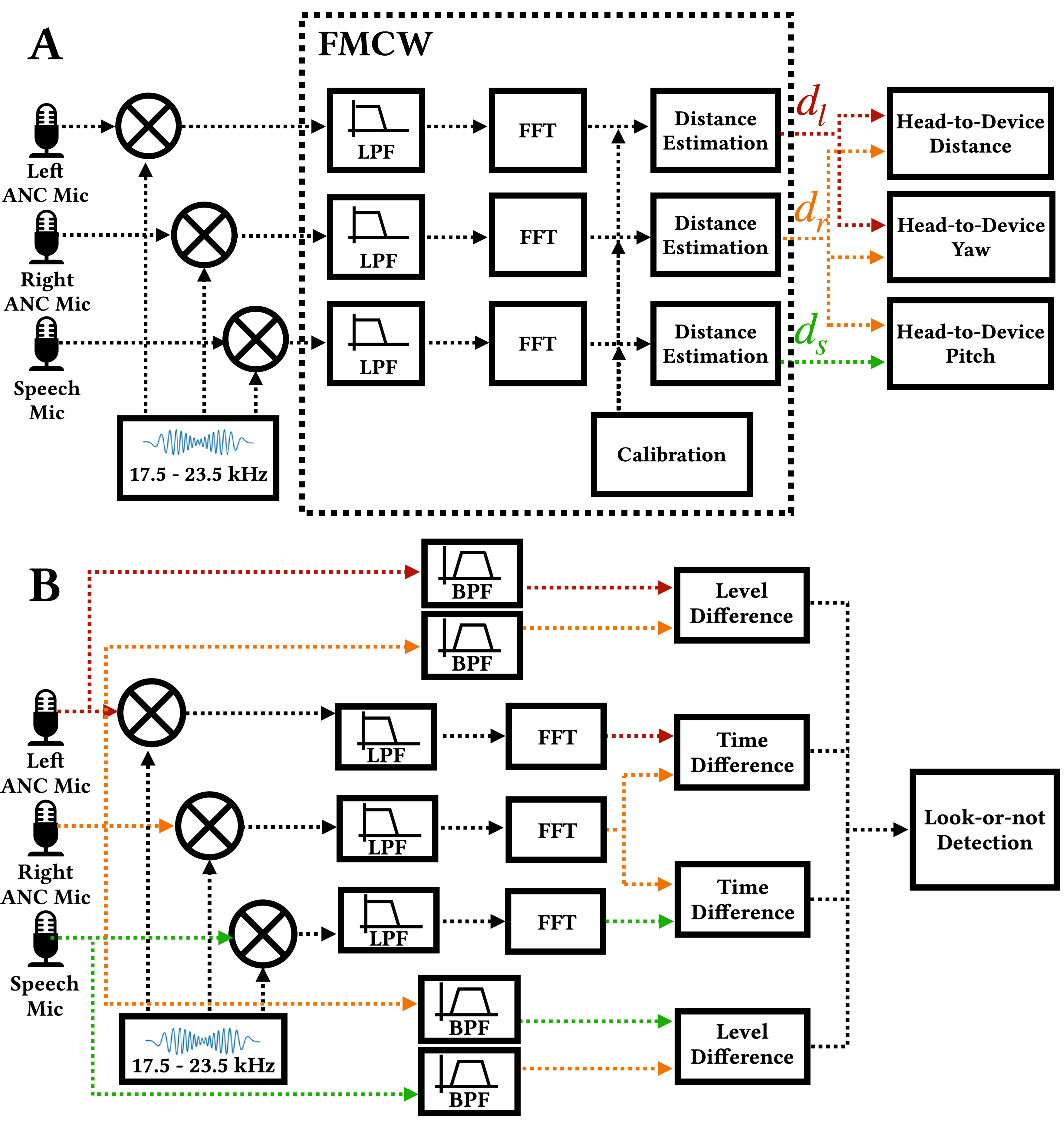}
  \caption{\change{\name{}~can enable continuous head position and orientation tracking (A) with FMCW-based acoustic ranging or binarized attention classification without calibration (B).} }
  \label{fig:method} 
\end{figure}

The time delay $t_d$ can be calculated with the frequency and phase from the mixed signal $y_m$. FMCW-based ranging methods with shared transmitter and receiver clock can directly calculate the 
peak at $f_p^d=\frac{B}{T}t_d$ in the frequency domain. From the peak frequency $f_p^d$, one can estimate the delay time and thus the distance.
However, with a common problem of distributed FMCW systems~\cite{Mao-MobiCom-CAT, Wang-2019-MilliSonic, DroneTrack}, \name{} has a separate transmitter (device speaker) and receiver (earphone microphone) with unsynchronized clocks. \change{Therefore, \name{} requires a calibration procedure to establish a reference position with the peak frequency of $f_p^0$, and the detail is described in Sec.\ref{sec:calibration}.}
The distance between the speaker and microphone can be calculated with the following equation, where $c$ is the speed of sound.

\begin{equation}
    D = c\frac{(f_p^d - f_p^0) T}{B}
    \label{equ:D}
\end{equation}

\subsubsection{Calibration}
\label{sec:calibration}

The calibration procedure is required for \textit{head tracking} but not mandatory for \textit{binarized attention detection}.  Referring to MilliSonic~\cite{Wang-2019-MilliSonic}, we require the user to place the left ANC microphone against the speaker with around a 2 mm gap for a couple of seconds (4 seconds are sufficient based on our evaluation in Sec.~\ref{subsec:config}). Therefore,  \name{} can 1) establish a reference position with the peak frequency of $f_p^0$; 2) perform an approximation synchronization by correlating the received signals with the original one; 3) handle the continuous clock time drift between the transmitter and the receiver by applying the linear curve fitting solution~\cite{Mao-MobiCom-CAT, Wang-2019-MilliSonic}. We acknowledge that this procedure limits \change{the convenience of using} \name{} in the real-world application. Alternative calibration methods are discussed in Sec.~\ref{subsection:dicuss_calibration}.

\subsubsection{Optimizations for missing LOS and the Doppler effect}
\label{sec:optimization}
 \change{Related acoustic ranging works~\cite{Mao-MobiCom-CAT,Wang-2019-MilliSonic, Cao-EarphoneTrack}} assume that there is a direct propagation path from the speaker to the microphone.
However, since the microphones are located on the side of the head, minimal head shifts can cause the microphones to be occluded from the speaker.
This loss of line-of-sight (LOS) significantly degrades the signal-to-noise ratio (SNR). It distorts the peaks ($f_p^d$) in the frequency domain, resulting in multiple peaks or the direct path peak merging with anomalous nearby peaks. Further, the quick head motion can introduce a significant Doppler effect. These issues can significantly degrade the tracking performance. Therefore, we developed and applied optimizations to the existing 
method to better support our application scenarios. 
Inspired by advanced techniques in the FMCW radar research field~\cite{merrill2001introduction,Liang2009,Chin-NLOS}, we extended CAT~\cite{Mao-MobiCom-CAT} with optimization approaches to solve the non-LOS and Doppler effect issues. \name{} adopts an inaudible triangular modulated chirp signal \change{to reduce the Doppler effect. Our implementation adopted} an up-chirp from 17.5 kHz ($f_0$) to 23.5 kHz ($f_1$) followed by a down-chirp to 17.5 kHz with a total sweep time of 42.7 ms (2048/48000). \name{} further averages the two parts of measurements at different edges of the triangular pattern~\cite{richards2014fundamentals}. Therefore, \name{} can achieve a more accurate distance estimation despite the frequency shift caused by the Doppler frequency, as prior works~\cite{merrill2001introduction, ReflecTrack} explained. To further increase the SNR, \name{} adopts a non-coherent integration method~\cite{Liang2009} by averaging the intermediate FFTs from a small set of recent frames (2 frames in our implementation). 

To get the correct peak corresponding to the direct path, we used the Constant False Alarm Rate (CFAR) adaptive thresholding algorithm~\cite{Weiss-cfar, richards2014fundamentals} on the FFT values of the mixed signal --- $y_m$. The algorithm combines the following heuristics: 1) an early and high peak that is closest to the previous peak is selected corresponding to the direct peak due to the continuous change in the distance; 2) when a sudden peak shift appear in one microphone channel but not the others, indicating a loss-track event, a fallback algorithm is utilized to predict the peak frequency from recent historical frames (5 frames in our implementation) through interpolation.

\subsection{Yaw and Pitch Estimation}
\label{sec:yaw_pitch}
The FMCW-based acoustic ranging technique provides three distances between the speaker and the earphones' three microphones. Two microphones used for active noise cancellation (ANC) sit at a similar elevation at the top of the earcup (see Fig.~\ref{fig:hardware}). A single speech microphone sits at a lower elevation on the right earcup. By comparing distances between the left and right ANC microphones from the speaker, yaw can be calculated. By comparing distances between the right ANC microphone and the speech microphone, the pitch can be calculated.

\begin{figure}
  \centering
  \includegraphics[width=\columnwidth]{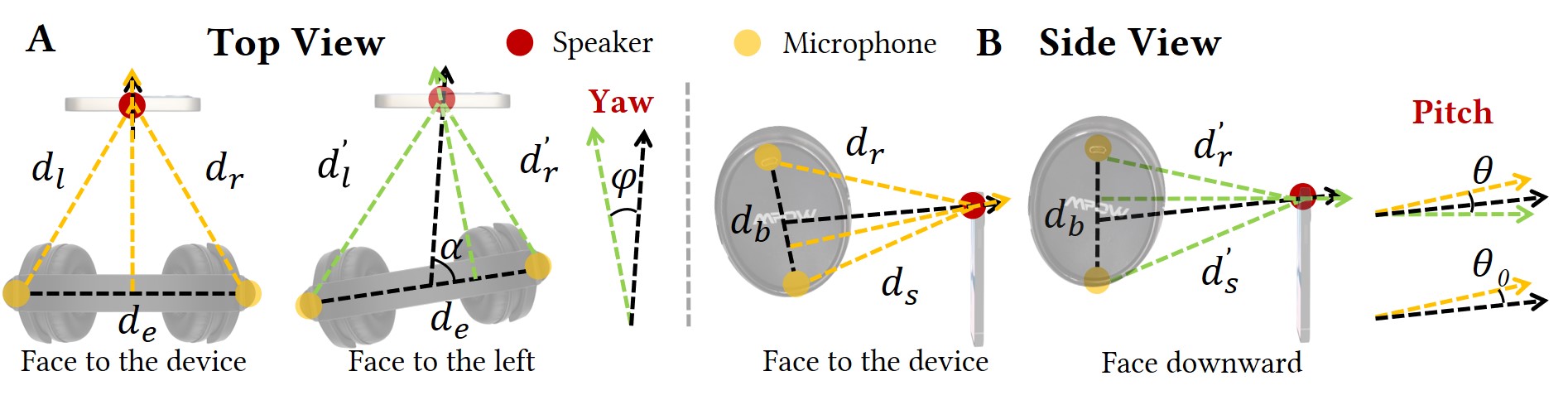}
  \caption{\name{}~estimates face orientation towards the sound source by calculating the angle between the median line and the altitude line. }
  \label{fig:angle} 
\end{figure}

The yaw and pitch angles are calculated as angles between the face orientation vector and the vector from the center of the head to the speaker location. The mic-speaker distances form triangles in the transverse (top view, see Fig.~\ref{fig:angle}A) and sagittal (side view, see Fig.~\ref{fig:angle}B) plane of the head. In each, the triangle's altitude is aligned with the face orientation vector, and the triangle's median is aligned with the vector between the head center and speaker. We refer to the distance from the speaker to the left ANC microphone as $d_{l}$, the right ANC microphone as $d_{r}$, and the right speech microphone as $d_{s}$. The distance between the left and right ANC microphones is $d_{e}$, which can be measured manually or set by an average value across users. The distance between the right speech microphone and the right ANC microphone is $d_{b}$, a known quantity. To explain our method, we will detail how the yaw angle is calculated. A similar approach is employed for pitch estimation. 
The length of the median line ($d_m$) is calculated as:
\begin{equation}
    d_{m} = \frac{\sqrt{2{d_{l}}^{2} + 2{d_{r}}^{2} - {d_{e}}^{2}}}{2}
    \label{equ:median}
\end{equation}
and the angle between the median line and the base line of the triangle ($\alpha$) defines yaw ($\varphi$) as follows:

\begin{equation}
    \alpha = \arccos(\frac{{d_{m}}^{2} + {\frac{1}{2}d_{e}}^{2}-{d_{r}}^{2}}{d_{m}d_{e}})
\end{equation}
\begin{equation}
\label{eq:yaw}
    \varphi = \alpha - 90^\circ
\end{equation}

The same approach can be used to calculate pitch ($\theta$), by replacing the triangle formed by $d_{r}$,$d_{s}$, and $d_{b}$ with the one formed by $d_{l}$,$d_{r}$, and $d_{e}$.
Therefore, we can obtain both the yaw ($\varphi$) and pitch ($\theta$) angles of the user's face towards the speaker. This is achieved by subtracting the current angle with the known initial angle ($\theta_0$), since the speech microphone is slightly skewed off in vertical from the right ANC microphone in the earphone design (Sec. \ref{sec:impl}).

\subsection{Binarized Attention Detection}
\label{sec:binary_detection}
\name{} requires a calibration procedure for continuous face tracking. However, we present an alternative binarized attention classification method that detects whether the user is looking at the device without any calibration. This binarized detection can still be useful in various scenarios, including attentive user interfaces~\cite{looktotalk, EyePliances,auralamp}. As Fig.\ref{fig:method}B shows, we first applied a bandpass filter with a frequency range of 17.5 kHz to 23.5 kHz to the audio signals from the three microphones. \change{We evenly divided the frequency range into 20 bands. In each band,} we can obtain the level difference (LD), which is the amplitude level ratio of audio signals from two microphones. We obtained 20 $\times$ 3 = 60 LD features among three microphones. We also adopted 3 time difference features between the microphones. \change{Each time difference feature is} the frequency gap between the peaks ($f_p^d$) in the frequency domain of the mixed signals \change{from two microphones}. Using the 63 features (Fig.~\ref{fig:method}B), \name{} can detect whether the user is looking at the device by training a binary classifier. In our implementation, we adopted the supported vector machine (SVM, RBF kernel, C = 1.0) as the binary classifier.

\section{Implementation}
\label{sec:impl}

\subsection{\name{} Hardware}
\label{subsec:hardware}
\subsubsection{Headphone Prototype}
\begin{figure}[ht]
  \centering
  \includegraphics[width=0.9\columnwidth]{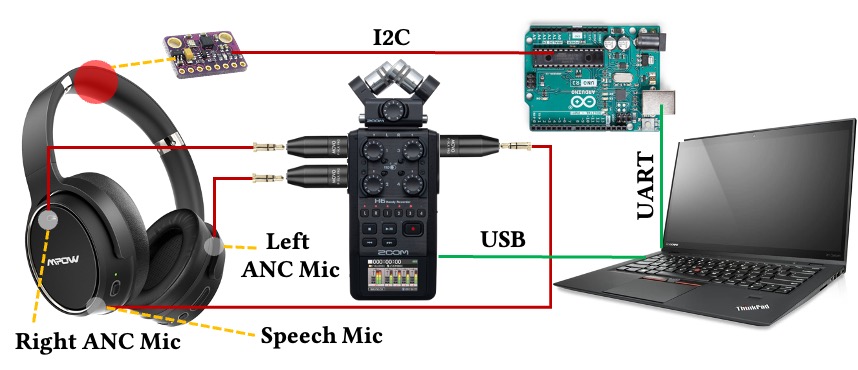}
  \caption{\name{}'s earphone hardware has a commodity earphone hardware (MPOW H19 for demonstration), an MPU-9250 IMU, an audio interface, and a laptop to process the audio signal.}
  \label{fig:hardware} 
\end{figure}

Modern ANC earphones share a similar design in the microphone placement~\cite{Rivera-Hybrid} as Fig.~\ref{fig:hardware} shows. Two microphones are located at the top of the earcups for collecting environmental noises. A speech microphone or microphone array sits at a lower elevation on one earcup. We adopted MPOW H19~\footnote{https://www.xmpow.com/products/mpow-h19-hybrid-noise-cancelling-headphones} \change{for evaluation. Further, we demonstrated \name{}'s applications using} Hush earphone by 233621~\footnote{https://www.233621.com/}, Sony WF-1000XM3~\footnote{https://www.sony.com.sg/electronics/truly-wireless/wf-1000xm3}, and ANC earbud --- Sony WH-1000XM3~\footnote{https://www.sony.com.sg/electronics/headband-headphones/wh-1000xm3} without an extra speech microphone. To obtain the high-resolution raw acoustic stream, we wired out two feed-forward ANC microphones and the speech microphone  with 3.5mm TRS plugins. The plugins were connected to a Zoom H6~\footnote{https://zoomcorp.com/en/us/handheld-recorders/handheld-recorders/h6-audio-recorder/} audio interface via VXLR to a 3.5 mm audio adapter. Zoom H6 supports up to 6 synchronized channels of real-time audio streaming through USB. Therefore, the audio signals from the three microphones on the earphone were streamed by the Zoom H6 to a Thinkpad X1 carbon laptop (CPU: i7-10710U, 6 cores, 1.1 GHz, RAM: 16GB, Storage: 512GB), which ran the audio signal processing algorithms in real-time. The sampling rate and the bit depth were set to 48 kHz and 16 bits. To further compare \name{}'s performance to the inertial measurement unit (IMU) based solution, we adopted the MPU-9250~\footnote{https://invensense.tdk.com/products/motion-tracking/9-axis/mpu-9250/} \change{IMU module}, which has a 3-axis accelerometer, a 3-axis gyroscope, and a 3-axis magnetometer. The data was streamed to the laptop with an Arduino Uno using the I2C protocol. The laptop read the IMU data with the same sampling rate --- 23.4 frames per second (fps). To avoid the effect of the speaker magnet, we mounted the IMU module to the top of the earphone. Before each measurement, we calibrated the magnetometer inside the IMU by drawing the $\infty$ shape in the air. 

\subsubsection{Audio Transmitter}
\label{subsec:mobile}
\change{A common} device with a speaker capable of generating inaudible ultrasonic sound (e.g., above 17 kHz) can be adopted as an audio transmitter. \change{During our evaluation}, a Samsung Galaxy S21 Ultra smartphone (256GB storage, 12GB RAM) with stereo speakers was adopted as the audio transmitter. Further, we demonstrated \name{}'s applications using Thinkpad X1 Carbon laptop ( Intel i7-10710U CPU, 16G RAM, 512G storage), Mi Watch (8GB storage, 1GB RAM), and Huawei Matepad PRO (10.8 inches, 256 GB storage). We generated a one-hour mono-channel audio file with continuous triangular chirp signals modulated signals (see Sec.~\ref{sec:optimization}). Then, the transmitter played the audio file from only one speaker using the HibyMusic~\footnote{https://store.hiby.com/} application, which supports the sampling rate and bit depth at 48 kHz and 16 bits, respectively.

\subsection{\name{} Software}
\label{subsec:implementation}
\begin{figure}[ht]
  \centering
  \includegraphics[width=0.9\columnwidth]{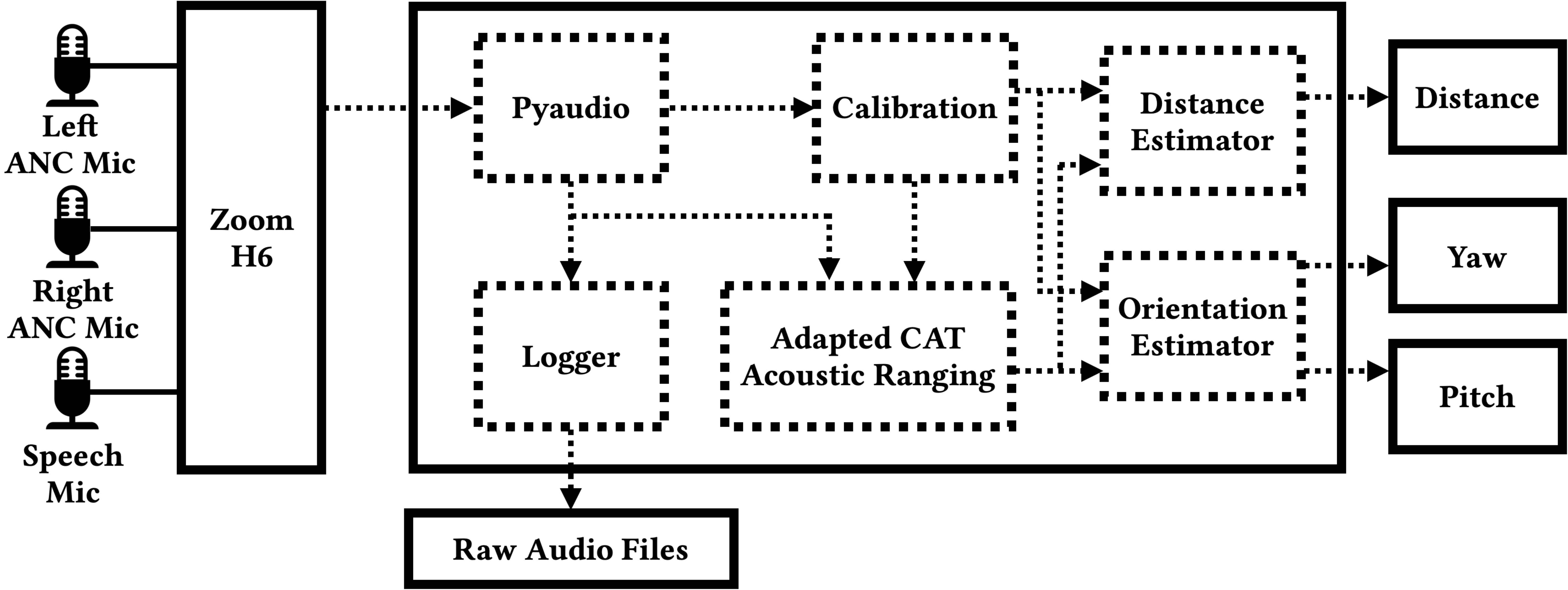}
  \caption{\change{\name{} software's key components.}}
  \label{fig:software} 
\end{figure}

We implemented \name{} (see Sec.~\ref{sec:distance}) using Python on the Thinkpad X1 Carbon laptop. As Fig.~\ref{fig:software} shows, we used PyAudio~\footnote{https://pypi.org/project/PyAudio/} to read the \change{triple-channel} raw audio signal from the Zoom H6 audio interface. All the raw audio data was stored for further offline analysis. The calibration module was launched when the user clicked the calibration button on the user interface. \name{} requires two parameters to be calibrated that are (1) the distance between the two ANC microphones ($d_e$) when the user wears the earphone (see Sec.~\ref{sec:yaw_pitch}), and (2) the reference origin for precise acoustic ranging (see Sec.~\ref{sec:calibration}). Then we pressed the \textit{Calibrate} button on the launch user interface and kept the two devices still for a 10-seconds duration. \change{Notably, 10 seconds are redundant for later evaluation (Sec.~\ref{subsec:config}).} Once the "Start" button is pressed, \name{} software displays the distance, yaw, and pitch values onto the launch user interface in real-time. Further, another interface popped up showing the measured distance curves with the three channels of the audio signal as Fig.~\ref{fig:lock}C shows.

\section{Evaluation}
\label{sec:study}
In this section, we describe the in-lab user evaluation study to benchmark \name{}'s performance of distance, orientation tracking, and binarized attention detection when the user is at different positions in relation to the device.

\begin{table*}[b]
  \centering
  \caption{The tracking performance within 9 grids. Each value group A/B/C indicates performance when the smartphone is placed on the height of 80/120/160 cm. Distance (mm),  Yaw ($^\circ$), Pitch ($^\circ$).}
  \small
    \begin{tabular}{|c|c|cc|cc|cc|}
    \hline
          \textbf{Y/X} &  & \multicolumn{2}{c|}{\textbf{-50 cm}} & \multicolumn{2}{c|}{\textbf{0 cm}} & \multicolumn{2}{c|}{\textbf{50 cm}}\\
    \hline
          & & \textbf{MedAE} & \textbf{IQR} &  \textbf{MedAE} & \textbf{IQR} &  \textbf{MedAE} & \textbf{IQR}\\
    \hline
     & Distance & 12.9/9.5/5.4 & 19.2/14.7/7.8 & 7.9/4.1/3.4 & 13.9/8.1/6.3 & 10.1/6.6/4.6 & 27.3/13.2/8.5  \\
    \textbf{50 cm} & Yaw  & 4.7/4.9/2.0 & 7.8/8.2/2.6 & 3.7/1.7/1.5 & 5.9/3.2/2.6 & 5.7/3.9/2.0 & 11.9/8.0/3.5  \\
    & Pitch  & 6.2/4.2/3.1 & 10.0/5.4/3.8 & 8.9/3.4/2.8 & 15.9/6.0/4.5 & 9.7/5.7/4.1 & 18.8/8.8/7.3  \\
    \hline
     & Distance  & 16.5/11.0/7.4 & 21.1/12.7/10.7 & 14.6/12.4/9.2 & 22.8/14.6/12.5 & 13.4/10.4/10.8 & 23.0/16.1/20.7  \\
     \textbf{100 cm} & Yaw   & 3.7/3.0/2.3 & 6.4/4.0/3.6 & 5.8/4.1/3.1 & 9.5/5.7/3.8 & 4.8/4.9/3.9 & 11.7/7.0/8.5  \\
    & Pitch & 6.3/4.5/3.6 & 9.5/6.0/5.4 & 6.7/6.6/4.9 & 12.5/9.1/6.7 & 8.4/6.5/7.4 & 13.5/9.8/13.9  \\
    \hline
     & Distance  & 24.5/17.4/14.6 & 29.8/19.6/25.0 & 19.4/11.2/10.6 & 55.1/17.7/22.3 & 19.6/15.5/13.1 & 41.5/29.5/23.8  \\
    \textbf{150 cm} & Yaw  & 4.7/3.7/4.0 & 10.8/6.4/6.7 & 6.6/5.0/3.9 & 14.9/8.9/6.9 & 5.5/6.6/4.5 & 12.0/13.4/11.5  \\
    & Pitch  & 9.3/7.3/6.7 & 15.0/11.9/10.7 & 9.8/8.3/6.3 & 15.1/12.2/10.3 & 8.3/7.5/8.2 & 13.6/11.9/14.4  \\
    \hline
    \end{tabular}%
  \label{tab:position_effect}%
\end{table*}%

\subsection{Participant and Apparatus}
We recruited 12 participants (7 females, 5 males) with an average age of 22.5 (SD = 3.4). Each had previously used earphones and smartphones. 
The experiment was conducted in a room with the size of 4 by 3 meters. To obtain the ground truth positions, we utilized the OptiTrack motion capture system (10 Prime 17 cameras) with its coordinate calibrated. The tracking markers were located at the phone's front speaker, which was located next to the front camera, and each microphone location on the earphone. In this evaluation study, we used a Samsung Galaxy S21 Ultra smartphone as our sound transmitter and MPOW H19 earphone as our receiver. A tripod with adjustable height was used to hold the phone at different heights.

\subsection{Experiment Design and Procedure}
\label{subsec:experiment}

Each participant was informed about the purpose and the procedure of the experiment. An experimenter assisted each participant in putting on the earphone and then measured the approximate distance between the left and right ANC microphones with a ruler. 
The experimenter conducted the calibration by placing the smartphone's speaker to the left ANC microphone of the earphone for 10 seconds.
To understand the effect of relative position on \name{}'s distance and orientation tracking accuracy, we created a 3D grid of test positions in front of the smartphone.
With the smartphone located at the origin (0,0) in the top view, three rows of grids were located at 50 cm, 100 cm, and 150 cm away from (0,0) in the $y$ direction. The three columns of grids were located at -50 cm, 0 cm, and 50 cm away from (0,0) in $x$.  We chose the maximum tracking distance to be around 160 cm --- (150, -50) or (150, 50) --- because we targeted \name{}’s usage scenarios within the range of a personal workspace. We tested three speaker heights: 80 cm, 120 cm, and 160 cm away from the floor. The participant adjusted the seated chair's height to a comfortable position. Therefore, the relative heights of the smartphone with respect to the earphone are different across the participants. During the user study, the lab noise levels ranged between 54.3 dBA to 62.7 dBA with a server running and people talking.

Each participant finished three head movement sessions at each 3D grid point. 
Each head movement session consisted of 6 sub-tasks: 1) look at the smartphone's speaker for 5 seconds, called the neutral state; 2) move forward and backward for 3 times; 3) \change{rotate} the head in the yaw direction for 3 times to the maximum range and return to the neutral state; 4) \change{tilt} the head in the pitch direction for 3 times to the maximum range and then return to the neutral state; 5) draw the zigzag shape from top left to the bottom right with 2 folds; and 6) randomly move the head for 3 seconds.
The order of the 2D grids was randomized under each height condition. We re-calibrated \name{} when we collected the data at a different height. Therefore, we conducted three calibrations in total. 
Each participant received a 20 USD gift card for their effort and time (40 minutes).

\subsection{Results}
\label{subsec:result}
\change{Same as CAT~\cite{Mao-MobiCom-CAT} and MilliSonic~\cite{Wang-2019-MilliSonic}, the deviation of \name{}'s measurements (distance, yaw, and pitch) follow non-Gaussian distributions, the median absolute error (MedAE) is a better measure compared to the mean absolute error (MAE). Therefore,} we report \name{}'s tracking performance using the MedAE and the interquartile range (IQR). Nonetheless, we also derive MAE in the discussion to compare against camera-based methods~\cite{Abate-2019-head-pose, Tan2018} in literature. The ground truth distance and orientation of a user's face towards the device were calculated as described in Sec.~\ref{sec:yaw_pitch} using coordinates of the tracker
attached to the microphones and the speaker in the OptiTrack system. We utilized Aligned Rank Transform Factorial ANOVA for within-subject non-parametric statistical analysis ($p < 0.05$) with Wilcoxon signed-rank test for post-hoc analysis ($p < 0.05$).

We summarize results in Table~\ref{tab:position_effect}. The table shows the MedAE and IQR of distance in millimeters, yaw in degrees, and pitch in degrees when the smartphone's speaker was placed at the height of 80, 120, and 160 cm. Each cell of the 3 $\times$ 3 cell represents one grid during the experiment (see Sec.~\ref{subsec:experiment}).

\subsubsection{Distance Tracking Accuracy}
\label{sec:distance_tracking}
\begin{figure}[ht]
  \centering
  \includegraphics[width=\columnwidth]{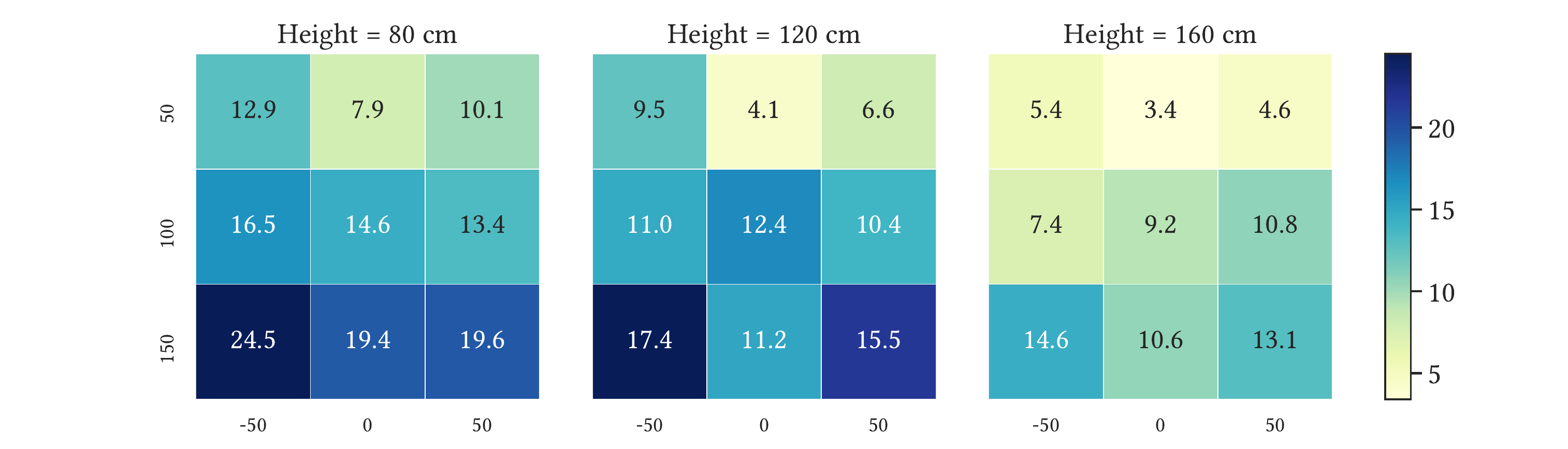}
  \caption{The median absolute error (mm) of the distance at different relative positions (cm) between the smartphone and the earphone.}
  \label{fig:result_distance} 
\end{figure}

Results show that \name{} can continuously track the distance from the user's head to the smartphone with a MedAE of 10.9 mm and an IQR of 18.8 mm. Statistical analysis shows that there are significant effects of 2D location (grid) ($F_{(8, 253100)} = 1260, p < 0.001$) and height ($F_{(2, 253106)} = 3115, p < 0.001$) on the distance tracking performance. Fig.~\ref{fig:result_distance} and post-hoc pairwise tests show that \name{} can achieve a better distance tracking performance when the user is closer to the smartphone ($p < 0.01$) and in the center column of grids ($p < 0.01$). Results show that \name{} can achieve the best performance when the smartphone was placed at height of 160 cm ($p < 0.001$). 

\subsubsection{Head Orientation Tracking Accuracy}
\label{sec:tracking}
Results show that \name{} can continuously track the user's head orientation with respect to the smartphone. The MedAE and IQR of the yaw angle were 3.7$^\circ$ and 6.8$^\circ$ and those of the pitch angle were 5.8$^\circ$ and 10.0$^\circ$. Results also show that the yaw range among all participants was from -81.1$^\circ$ (s.d. = 10.1) to 76.9$^\circ$ (s.d. = 9.1), the pitch range was from -87.8$^\circ$ (s.d. = 6.3) to 84.3$^\circ$ (s.d. = 9.9). \change{The maximum, MedAE, and IQR of head orientation speed of the yaw angle were 250.9 $^\circ$/s, 185.6 $^\circ$/s, and 33.2 $^\circ$/s and those of the pitch angle were 185.6 $^\circ$/s, 13.6 $^\circ$/s, and 44.1 $^\circ$/s.} These results can be helpful references for experience development.

\begin{figure}[ht]
  \centering
  \includegraphics[width=\columnwidth]{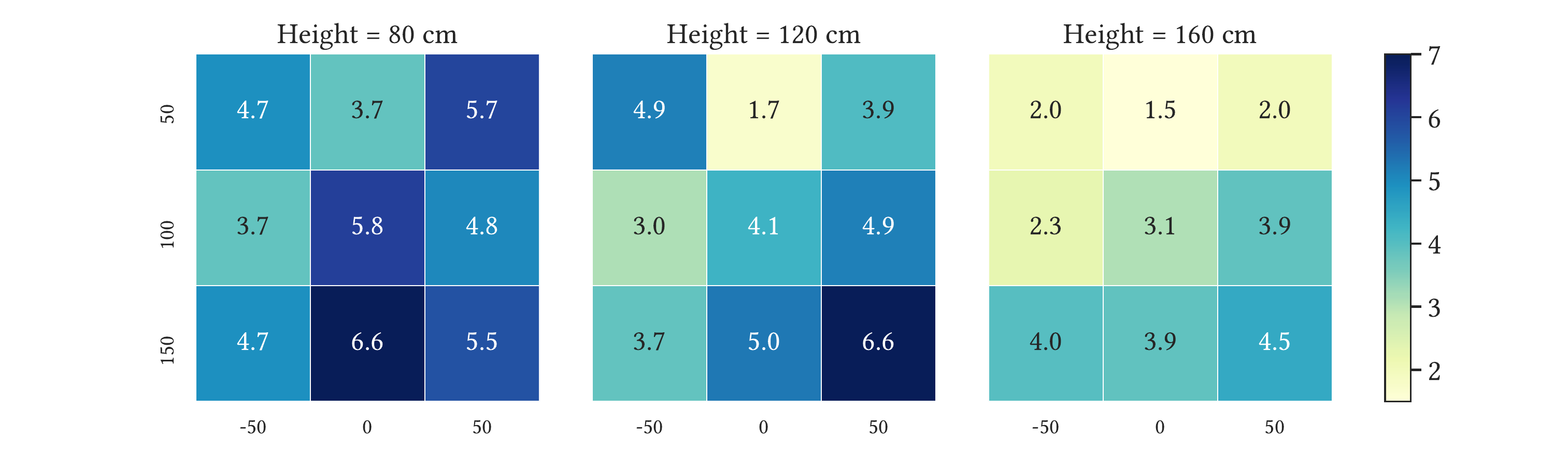}
  \includegraphics[width=\columnwidth]{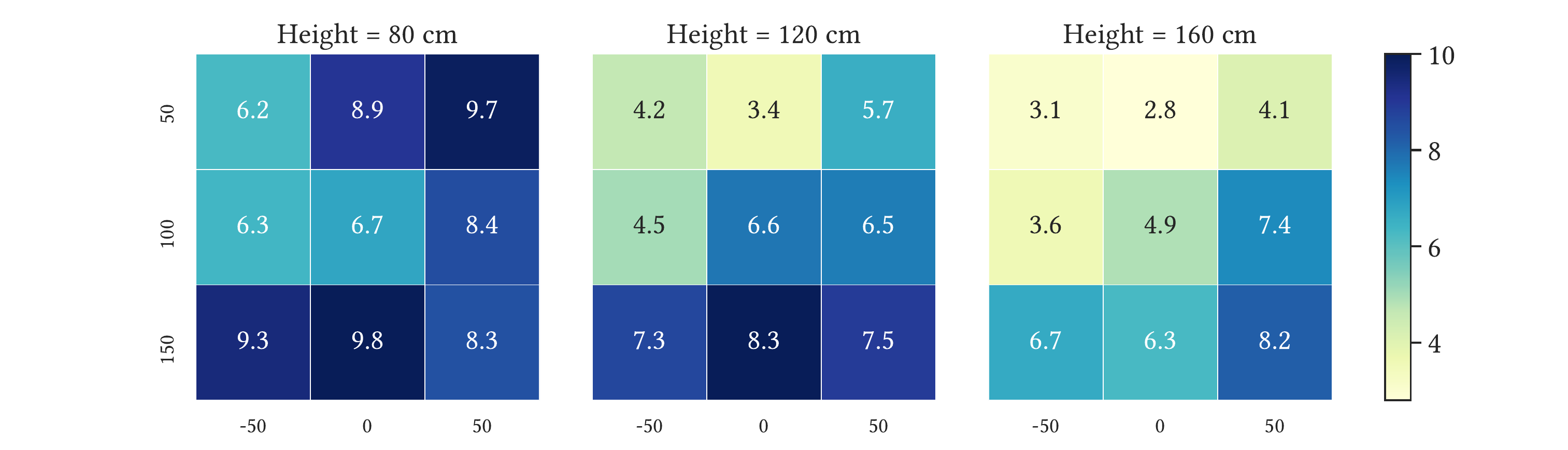}
  \caption{The median absolute errors of the yaw (left) and pitch (right) angles with different relative position (cm) between the smartphone and the earphone. }
  \label{fig:result_orientation} 
\end{figure}

We further evaluated the effect of the relative position of the user's head with respect to the smartphone on the orientation tracking performance. Statistical analysis shows that there \change{are} significant effect\change{s} of 2D location (grid) on the yaw ($F_{(8, 234813)} = 745, p < 0.001$) and the pitch ($F_{(8, 226907)} = 730, p < 0.001$) tracking performance. Fig.~\ref{fig:result_orientation} and post-hoc pairwise tests show that \name{} can achieve a better tracking performance when the user is closer to the smartphone and in the center column of grids ($p < 0.01$) in general. Further, there \change{are} significant effect\change{s} of height on the yaw ($F_{(2, 234819)} = 1056, p < 0.001$) and the pitch tracking performance ($F_{(2, 226913)} = 390, p < 0.001$). Again, \name{} can achieve a better performance when the smartphone was placed at height of 160 cm ($p < 0.05$) as compared to the other heights. Further, we observed a significant effect of the relative height ($p < .01$) of the earphone to the smartphone's speaker but not the absolute sitting height of the participant ($p = 0.07$) on the tracking performance.

\subsubsection {Binarized Attention Classification Accuracy}
Using the trackers on the smartphone, we labeled the smartphone's margin as a rectangle in the OptiTrack coordinate space. We regarded the ground truth of orienting to the device as the vector of the user's head intersecting with this rectangle. When evaluated \change{on} the performance through \textit{leave-one-out cross-user validation} for the look-or-not binary classification, \name{} can achieve an average classification accuracy of 93.5\% across users (
s.d.= 2.5\%). The calculation latency is within 42 ms. 

\subsubsection{Effect of Setup Configuration}
\label{subsec:config}
During the user study, the experimenter measured the distance between two ANC microphones ($d_e$ in Fig.~\ref{fig:angle}) manually. Results show that the mean absolute error of the measurement with a ruler is 3.4 mm against OptiTrack. \change{But choosing which one from these two measuring methods has} no significant effect (p = 0.33) on the yaw tracking performance. Further, when setting $d_e$ to an fixed distance of 235 mm (considering human's head breadth 155 mm~\footnote{https://en.wikipedia.org/wiki/Human\_head} + extra earcup depth 40 mm $\times$ 2), \name{} was still able to achieve a satisfying performance that the MedAE in yaw direction increases by only 1\% without a significant difference ($p = 0.1$). Therefore, \change{there is no evidence that it is necessary to manually measure the distance between two ANC microphones.}

We applied a redundant clock-sync calibration duration --- 10 seconds. Since we recorded all the data from the study, we reran our method with a 4-second calibration duration, resulting in only a 3\% increment in the median absolute error across all angles and distances, adequate for a whole experiment session of ~15 minutes.

Our evaluation environment had a background noise from 54.3 dBA to 62.7 dBA with a server running and people talking, indicating the robustness of ultrasonic ranging against noise, which aligns with the result from MilliSonic~\cite{Wang-2019-MilliSonic}. 

\subsubsection{Comparison with CAT Baseline Method}
\label{sec:lose_track}

\begin{figure}[ht]
    \centering
    \includegraphics[width=0.85\columnwidth]{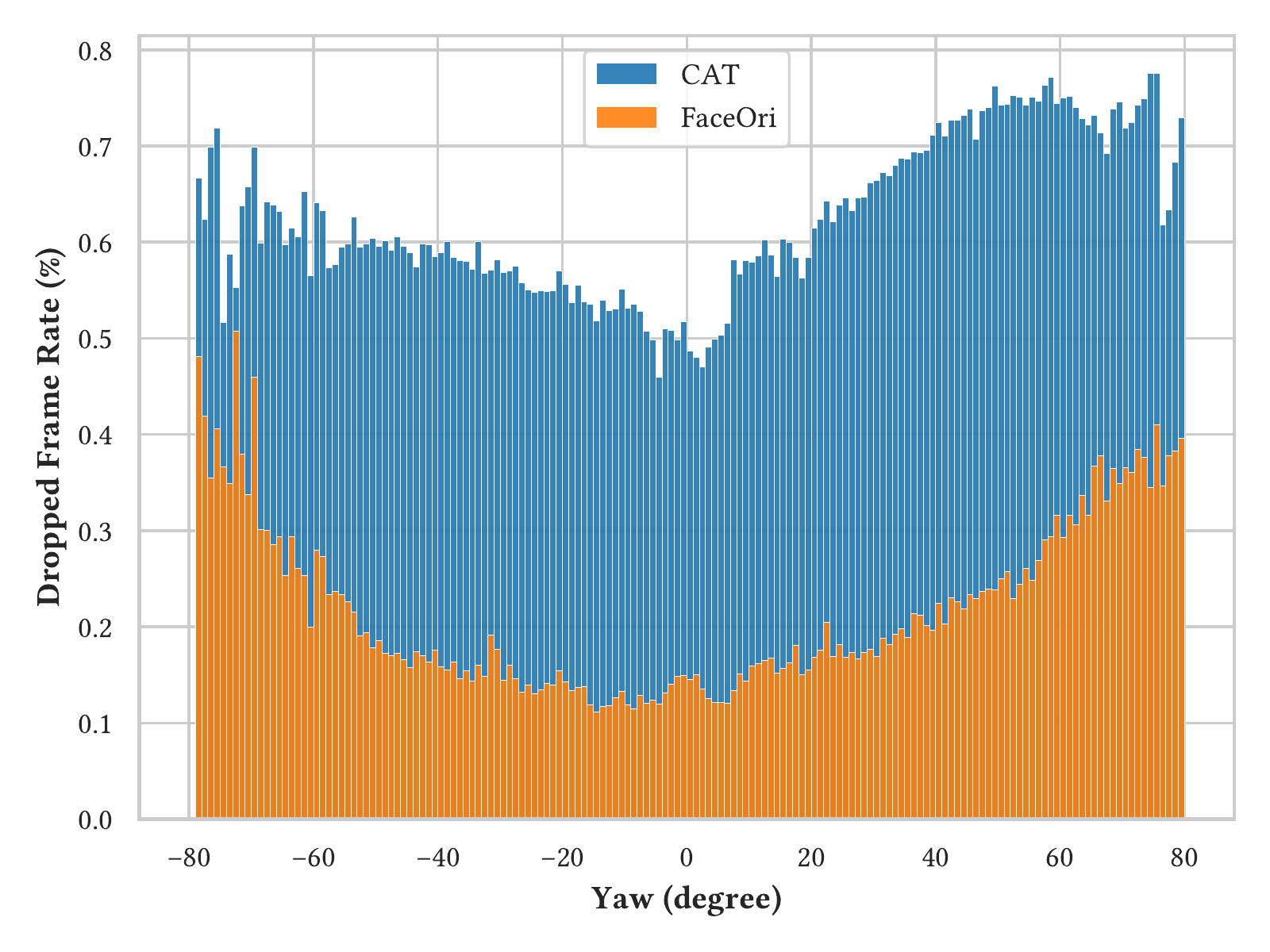}
    \includegraphics[width=0.85\columnwidth]{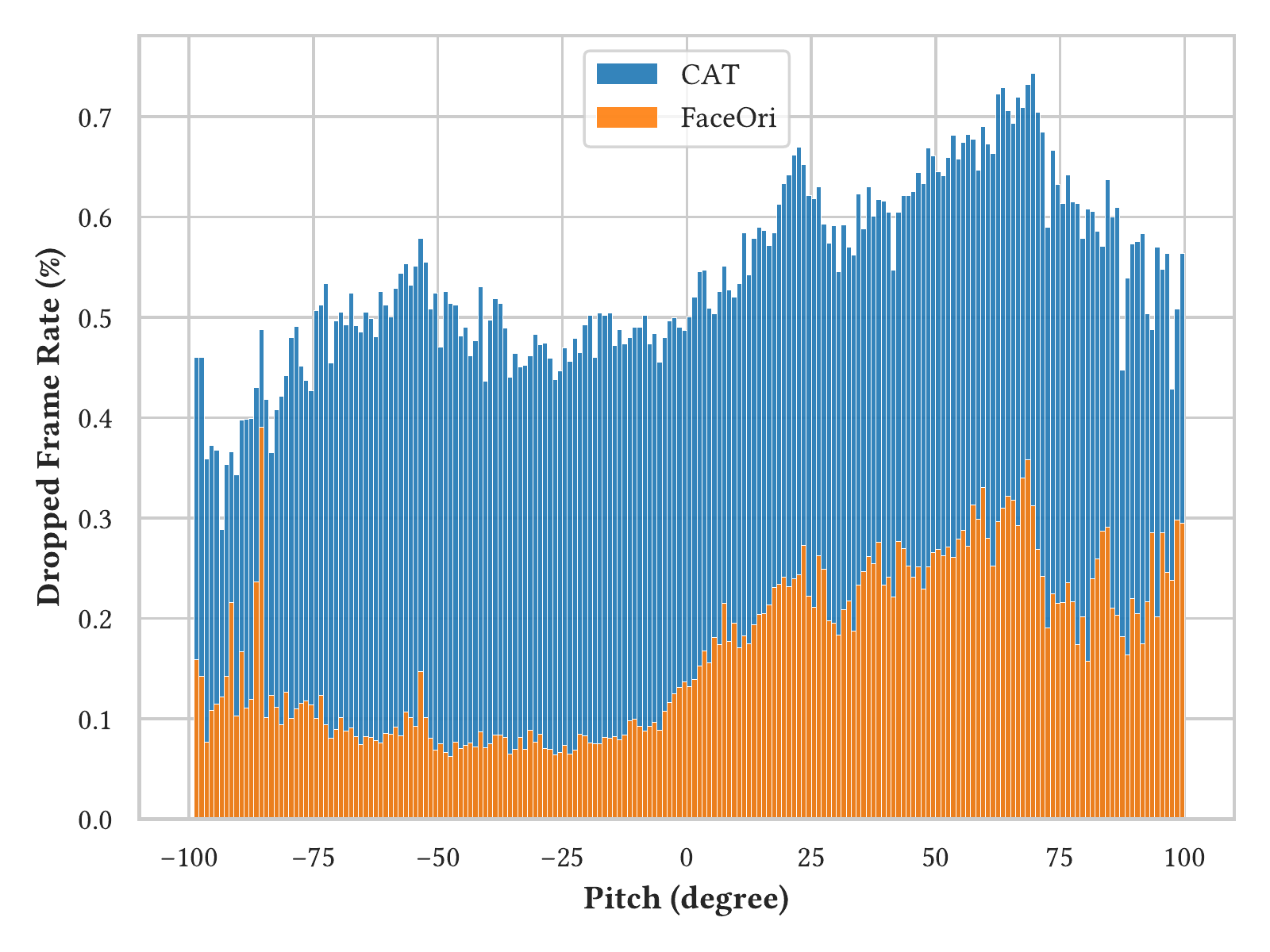}
    \caption{\name{} has less dropped frames compared with CAT acoustic tracking method~\cite{Mao-MobiCom-CAT}.}
    \label{fig:lose_track}
\end{figure}

We evaluated the effectiveness of our optimization method mentioned in Sec.~\ref{sec:optimization} to overcome the issues introduced by the head motion. We compared our method to the baseline CAT acoustic ranging method~\cite{Mao-MobiCom-CAT}.
Results show that our method (10.9 mm in the distance, 3.7° in yaw, and 5.8° in pitch) significantly outperforms CAT (42.0 mm, 11.0° and 11.6°) on our collected dataset ($p < 0.001$ for all cases).
Here, we define a dropped frame with the feature of a large distance offset away from the ground truth (37.6 mm as our threshold - $2.0 \times IQR$). Fig.~\ref{fig:lose_track} shows the dropped frame rate along with the ground-truth yaw (left figure) and the pitch (right figure) angle. 
Results show that our method can effectively decrease the dropped frame rate with an average rate of 14.8\% versus 52.8\% (CAT method). \change{Head rotation of a larger amplitude} in yaw or pitch angles results in more dropped frames.
This demonstrates that our method is more robust against the \change{non-LOS} and Doppler effect introduced by the head motion. Further, the large spread of the head orientation angle and speed also indicate \name{}'s robustness against noises introduced by the head motions. 

\subsubsection{Comparison with IMU-Based Head Orientation Tracking Solutions}
To compare the performance of \name{} and the IMU-based solution (MPU-9250 with Arduino code~\footnote{https://github.com/hideakitai/MPU9250}), we performed calibration to the IMU by aligning its initial yaw and pitch angle with the OptiTrack coordinate system. Results show that IMU-based solution can continuously track the user's head orientation with a MedAE of 17.2$^\circ$ (IQR = 330.4) in yaw, and 4.9$^\circ$ (IQR = 10.8) in pitch. The pitch tracking performance is significantly better than \name{} ($p < 0.001$). However, we observed significant yaw drift, a well-studied problem in IMU tracking~\cite{LaValle-Oculus}, even with the reference calibration provided by the magnetometer. Therefore, the IMU-based solution is insufficient for our applications, which require more accurate yaw estimation. 

There are additional functionality limitations regarding \change{to} the IMU-based solution. \change{The} IMU tracks \change{it orientation} relative to the inertial world reference frame rather than the mobile device reference frame as \name{} does. Modern earphones with IMUs do not contain the magnetometer due to the speaker's strong magnet. Further, IMU cannot provide accurate absolute distance to the mobile device (>20 cm error with calibration~\cite{Xu-IMU-2019}).

\subsubsection{Comparison with Camera-Based Head Orientation Tracking Solutions}
To compare with camera-based solutions in the literature, we measure \name{}'s performance with a mean absolute error of 8.3$^\circ$ in yaw and 9.6$^\circ$ in pitch. This is comparable with cutting-edge RGB camera-based technique, which can track head orientation with MAE of 7.6$^\circ$ in both yaw and pitch~\cite{Abate-2019-head-pose}. Further, RGBD-based methods (ARKit) achieved higher performance  --- 1.8 mm in the distance, 0.9$^\circ$ in yaw, and 0.7$^\circ$ in pitch~\cite{Tan2018}. However, \name{} has advantages in a wider field of view, preserving visual privacy, and has the potential to support the interaction with devices without cameras (e.g., smartwatch). \name{} can be complementary to the vision-based method regarding usage scenario, power consumption, privacy, etc. 

\subsubsection{Sensor Fusion of \name{} and IMU} \change{To further reduce the power consumption} during real-world deployment, we can adopt a sensor fusion method by combining \name{} and the IMU if available. \name{} can calibrate the IMU every certain amount of time $T_{cal}$. Therefore, we can track the yaw and pitch angles with higher accuracy than the IMU-based solution but consume less power than the sole ultrasonic-ranging-based solution. Since the IMU already achieved a better tracking performance in pitch than \name{}, we evaluated the sensor fusion method on the yaw angle. We ran the \name{} for 0.5 seconds to establish an accurate yaw angle and distance. Then we tested the effect of $T_{cal}$ (in second) on the tracking performance in yaw. Results show that the sensor fusion method can achieve a MedAE of 5.1$^\circ$ (IQR = 14.2$^\circ$), 7.0$^\circ$ (IQR = 20.0$^\circ$), and 9.7$^\circ$ (IQR = 21.5$^\circ$) when $T_{cal} = 1, 3, 5$ seconds. 
\section{Applications}
\label{sec:app}
\name{}\ provides three outputs as a user interacts with a particular computing device: \textit{distance measurement}, \textit{binarized attention detection}, and \textit{continuous orientation tracking}. These metrics can be used individually or in conjunction to enable and enhance applications.

\subsection{Activity and Gesture Sensing}
\label{subsec:exc}
\begin{figure}[h!]
    \centering
    \includegraphics[width=0.9\columnwidth]{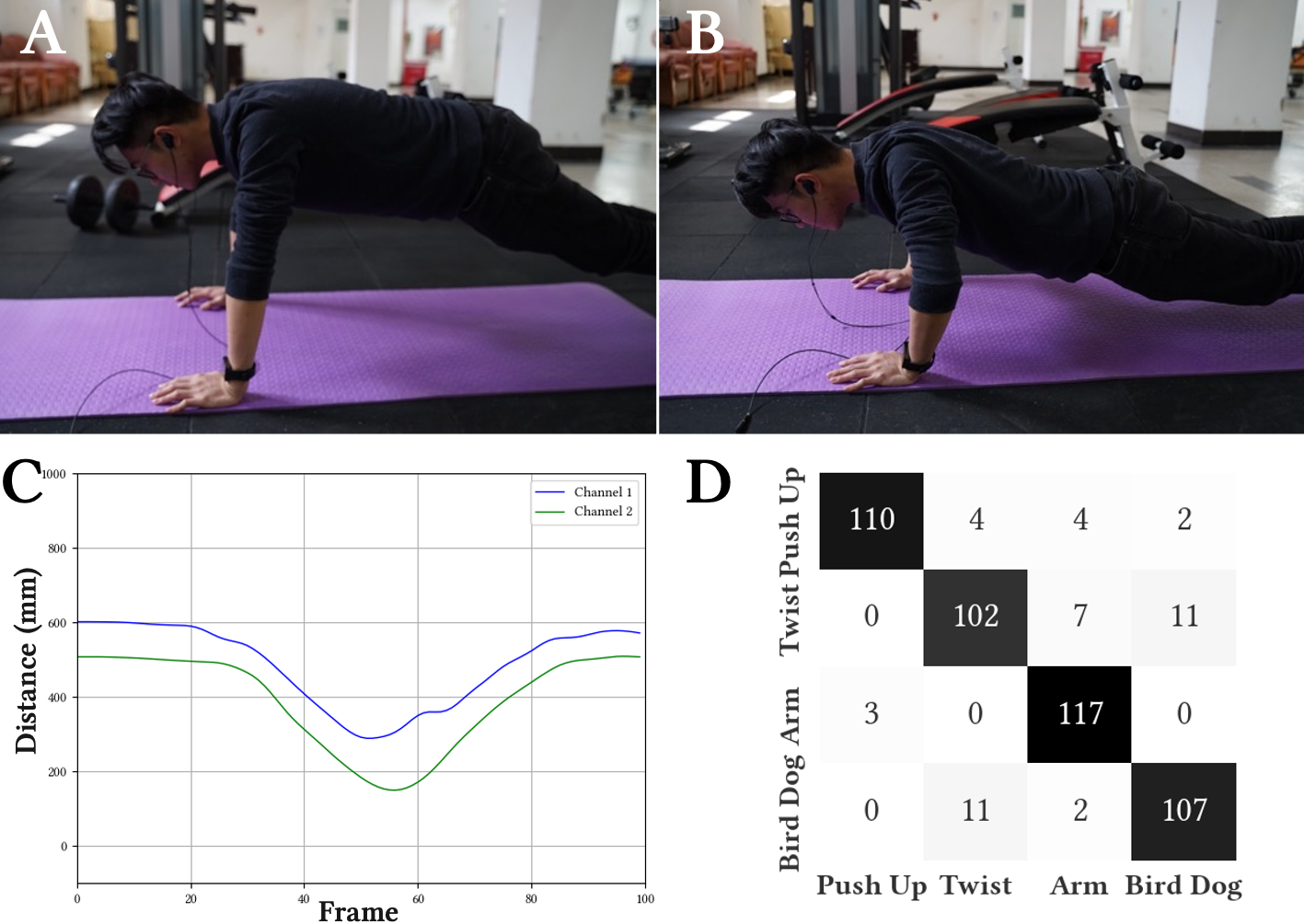}
    \caption{\change{\name{} can track and count the exercise activities with a smartwatch as the audio transmitter and an ANC earbud as the receiver. (A) (B) User does push-ups. (C) The distances between two ears and the speaker when user does a push-up. (D) Confusion matrix of the four classifications (push-up, body twist, touching shoulder, and bird dog). }}
    \label{fig:exercise}
\end{figure}

The head distance and orientation measurements can be used to analyze user activity or capture explicit user input. We prototyped an exercise smartwatch application that can count the number of exercise repetitions \change{that} a user performs by analyzing the periodicity of the \textit{distance measurement} provided by \name{} (Fig. \ref{fig:exercise}).
We used the Sony WH-1000XM3 earbud as the receiver and the Mi Watch as the transmitter in this application. 

To evaluate the feasibility of \name{} in activity recognition, we conducted a recognition evaluation for this application. Participants were asked to perform four activities: push-up, body twist, touching shoulder with contralateral hand (arm for abbreviation), and bird dog for 3 rounds with 5 repetitions per round. We manually segmented the dual-channel raw audio signal and aligned data to the length of 160 by interpolation. We chose 7 features for each frame: the distance between two microphones and smartwatch, the first derivative of two distances, the level difference between two microphones, and whether the signals of two channels lose track. Using SVM (RBF kernel, C = 1.0) to classify each exercise activity, \name{} can achieve an average accuracy of 90.9\% using leave-one-out cross-user validation.

The \textit{continuous orientation tracking} metric could also be used to enable gesture input. By analyzing oscillations in pitch and yaw, "yes" and "no" head shake gestures can be recognized. Finally, \textit{continuous orientation tracking} could drive a selector or pointer for accessible interfaces where a user may lack muscle control below the neck.

\subsection{Context-aware and Attentive User Interfaces}
\label{sec:attentive_UI}
\begin{figure}[h!]
    \centering
    \includegraphics[width=\columnwidth]{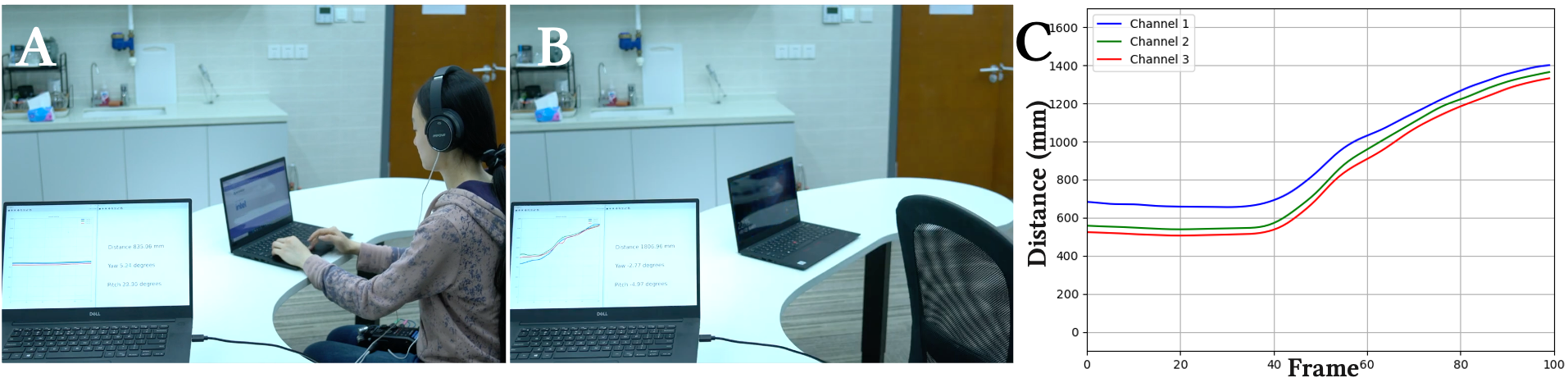}
    \caption{\change{\name{} dims the device when the user moves beyond a certain distance threshold away from the device. (A)(B) The User walks away from the laptop and the screen is dimmed. (C) \name{} measures distances from the laptop speaker to the three microphones in the earphone.}}
    \label{fig:lock}
\end{figure}

Real-time data on user position
can be used to drive smarter, more \change{context-}aware interfaces. \textit{Distance measurement} can be used to lock a phone or laptop or dim the screen when the user moves beyond a certain distance threshold away from the device (Fig.~\ref{fig:lock}). We implemented this example on a Thinkpad X1 Carbon laptop. The distance threshold was set to 1.5 meters. 

\begin{figure}[h!]
    \centering
    \includegraphics[width=\columnwidth]{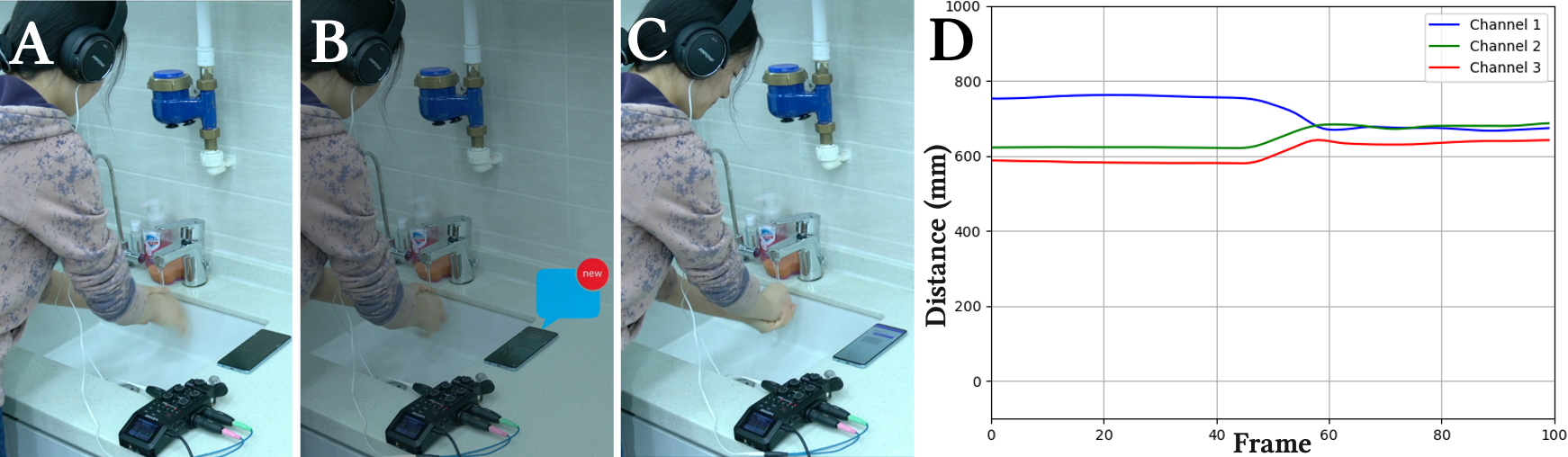}
    \caption{\change{\name{} can light up the smartphone and open to the message application when the user is unable to easily perform touch input. (A) The user is washing her hands. (B) When a message comes, (C) the user turns her head towards the phone to wake it up, and then the detailed message is displayed. (D) \name{} measures distances from the smartphone speaker to the three microphones in the earphone.}}
    \label{fig:wet_hand}
\end{figure}

\textit{Binarized attention detection} can help ease switching between multiple tasks or points of interest. For example, as a user follows a recipe video on their laptop, the video can automatically pause as the user turns to the stove or cutting board and resume when they return their attention to the screen. In Fig.~\ref{fig:wet_hand}, a user can provide input to their smartphone device even if they are otherwise preoccupied and unable to easily \change{perform} touch input.

\subsection{Attentive Detection from Multiple Devices}
\label{sec:multi-devices}
\begin{figure}[h!]
    \centering
    \includegraphics[width=0.9\columnwidth]{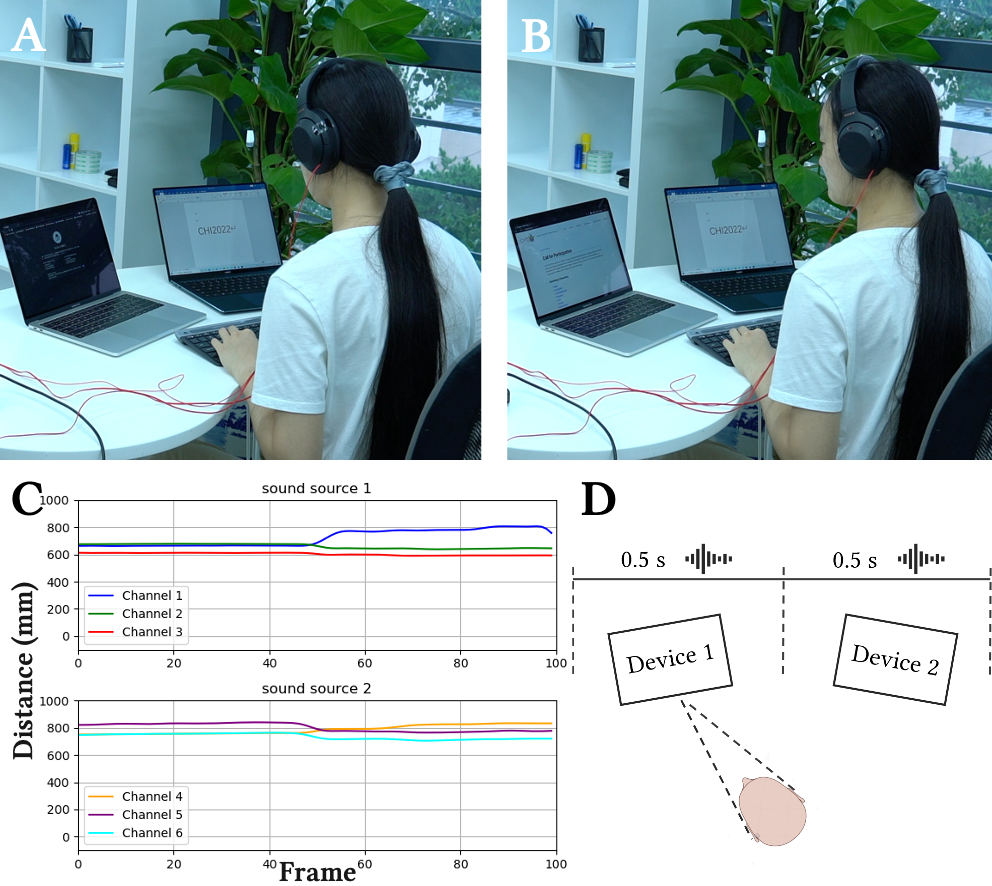}
    \caption{\change{The mouse and keyboard can pair to the device that the user faces towards automatically. (A)(B) The user switched her attention between two laptops. (C) \name{} measures distances from each laptop speaker to the three microphones in the earphone. (D) The time-multiplex approach for the multi-device application.}}
    \label{fig:multi_device}
\end{figure}

\name\ can be used to detect when users direct their intention toward a specific device. By orienting toward a smart speaker, a user can issue a command without requiring a keyword.
Finally, we prototype a demonstrative application that applies \name's \textit{binarized attention detection} in a simple multi-device scenario. As a user looks between \change{two laptops}, the keyboard and the mouse pair automatically to the laptop that the user watches (Fig. \ref{fig:multi_device}). To implement this application, the devices share a central server and time-multiplex their transmitted chirps.

We implemented the multi-device application on two Thinkpad X1 Carbon laptops. We used a third laptop as the proxy to 1) transfer the mouse and keyboard inputs to the two Thinkpad laptops; 2) run the \name{} algorithm to recognize which device the user orients to. These three laptops were connected through WiFi. The proxy coordinated the two Thinkpad laptops to let them emit the ultrasonic chirp signal alternately, 0.5 seconds for each. Therefore, the proxy knew which device was emitting the sound and then performed the \textit{binarized attention detection} to detect which device the user orients to. The first two seconds are skipped due to the delay of speakers and echoes. The time-multiplex approach caused a delay during recognition, but there was no evidence that it would influence the tracking performance.

\section{Discussion and Future Work}
\label{sec:disc}
This paper proposes \name{}, a novel \change{end-to-end} head position and orientation tracking \change{system} based on acoustic ranging using existing microphones in commodity earphones. Due to its high tracking performance, \name{} can support a wide range of novel interaction applications.
In this section, we discuss the findings, limitations, and avenues for future work. 

\subsection{Alternative Calibration Methods}
\label{subsection:dicuss_calibration}
The biggest limitation of \name{} is the requirement of calibration (Sec.~\ref{sec:calibration}). During our evaluation, to enable accurate continuous acoustic ranging, \name{} synchronizes the transmitter and receivers by holding one of the microphones to the speaker \change{every session}. However, existing work found that the calibration is only required once in each battery circle~\cite{Cao-EarphoneTrack}. 
Therefore, the per-session calibration is not necessary. We would expect future work to validate a per-battery-circle calibration method. 

To make the calibration procedure more user friendly, future work could explore using the front-facing RGB or RGBD camera on a device (if available) to establish the reference point and synchronize the clocks of the \change{device} and the wireless earphone. Further, calibration can be completely side-stepped if the earphone is connected to the transmitter via Bluetooth 5.0 or other wireless channels method with time synchronization protocol; thus, a sufficiently synchronized clock can be established. Future work can explore combining ultrasonic ranging with synchronization provided over these channels, avoiding calibration and additional complexity for drift compensation. A heuristic can be applied for a quick calibration procedure for applications requiring absolute distance but not requiring high accuracy. For example, the user can be instructed to hold their phone out at arm's length, and the origin can be set by substituting the average human arm length.

Notably, binarized attention detection requires no calibration \change{and} can be useful in various applications. Further, applications that use only a relative distance (such as the exercise application in Fig. \ref{fig:exercise}) do not necessarily require calibration. 


\subsection{Deployment and Generalizability}
\label{sec:real-world}
We developed and evaluated \name{} using the existing hardware in commodity earphones \change{and mobilephones}. To clarify, our test hardware is a proof-of-concept to evaluate our end-to-end face orientation and distance tracking system. In our implementation, we wired the built-in ANC microphones to an off-board laptop to host the signal processing program. However, modern ANC earphones have on-device processors. For instance, Sony 1000XM3 has a CSR8675 chip with DSP (120 MHz, 48kHz audio sampling rate) and MCU (80 MHz). We believe the proposed algorithms can be deployed on the micro-controller in the future \change{for real-world applications}.

We evaluated \name{}'s performance with only one type of earphone and one type of smartphone. However, we observed that ANC earphones adopt a common design in microphone placement as the tested one. Further, many newer models of earphones possess an extensive array of distributed microphones (e.g., Apple AirPods Max and the Bose 700). These characteristics can further improve the performance and increase the degrees of freedom. We expect future work to investigate \name{}'s generalizability across earphone models.

To further improve robustness and performance for real-world deployment, we would expect future work to further evaluate the effect of ambient noise on \name{}'s performance in various mobile scenarios. Meanwhile, future work can explore a more comprehensive sensor fusion method using the absolute (in device-frame) orientation provided by \name{} with the relative (with respect to an inertial reference frame) information supplied by the IMU.

\subsection{Supporting Multiple Devices}
\label{subsec:multi-device}
In section~\ref{sec:attentive_UI}, we briefly demonstrated a possible time-multiplexed solution to support multiple audio transmission devices, allowing \name{} to enable richer multi-device applications. However, this method assumes that all transmitters and the receivers can communicate via an additional channel (e.g., WiFi). We expect future work to explore other audio-only solutions to enable multiple device applications, such as using a frequency or phase-modulated chirp signal to provide unique device identification. Further, future work can adopt existing wireless multi-transmitter communication methods such as frequency hopping or code-multiplexing.

\section{Conclusion}
\label{sec:conclusion}

In this work, we have presented \name{}, a novel spatial input technique using ultrasonic ranging.  \name{} leverages the microphones found in typical active noise cancellation (ANC) earphones to glean user head proximity and orientation with respect to a computing device that emit an inaudible chirp from its speaker. Through a user study, we evaluated  \name{}'s performance for continuous head position and orientation tracking, and binarized attention detection. We explored and demonstrated how \name{} can be used to capture user activity and gestural input, \del{make applications more immersive }and enable more context-aware interactions. As the number and type of computing devices continue to proliferate, techniques like \name{} can help to make our interaction experiences more human-centered.

\begin{acks}
This work is supported by the Natural Science Foundation of China (NSFC) under Grant No.6213000120, 62002198, Tsinghua University Initiative Scientific Research Program, the China Postdoctoral Science Foundation	under Grant No.2021M691788, 
Beijing Key Lab of Networked Multimedia, and the Institute for Guo Qiang and Institute for Artifcial Intelligence, Tsinghua University.

\end{acks}

\bibliographystyle{ACM-Reference-Format}
\balance
\bibliography{ref}

\end{document}